# Improving Emotion Recognition Accuracy with Personalized Clustering

**Laura Gutiérrez-Martín[1,*], Celia López Ongil[1,2], Jose M. Lanza-Gutiérrez[3], and Jose A. Miranda Calero[4]**

[1] Department of Electronics, Universidad Carlos III de Madrid (UC3M), Spain
[2] Gender Studies Institute, Universidad Carlos III de Madrid (UC3M), Spain
[3] Department of Computer Science, Universidad de Alcalá (UAH) Spain
[4] Embedded Systems Laboratory, École Polytechnique Fédérale de Lausanne (EPFL), Switzerland
[*] corresponding author(s): Laura Gutiérrez-Martín (lagutier@ing.uc3m.es)

**Abstract.**

Emotion recognition through artificial intelligence and smart sensing of physical and physiological signals (Affective Computing) is achieving very interesting results in terms of accuracy, inference times, and user-independent models. In this sense, there are applications related to the safety and well-being of people (sexual aggressions, gender-based violence, children and elderly abuse, mental health, etc.) that require even more improvements. Emotion detection should be done with fast, discrete, and non-luxurious systems working in real-time and real life (wearable devices, wireless communications, battery-powered). Furthermore, emotional reactions to violence are not equal in all people. Then, large general models cannot be applied to a multiuser system for people protection, and customized and simple AI models would be welcomed by health and social workers and law enforcement agents. These customized models will be applicable to clusters of subjects sharing similarities in their emotional reactions to external stimuli. This customization requires several steps: creating clusters of subjects with similar behaviors, creating AI models for every cluster, continually updating these models with new data, and enrolling new subjects in clusters when required. A methodology for clustering data compiled (physical and physiological data, together with emotional labels) is presented in this work, as well as the method for including new subjects once the AI model is generated. Experimental results demonstrate an improvement of 4% in accuracy and 3% in f1-score w.r.t. the general model, along with a 14% reduction in variability.

I. Introduction.

Recently, Affective Computing (research on people's emotional reactions to certain external stimuli, and how these reactions can be measured and scaled using different types of sensors, [1]) is achieving important results in terms of emotion detection with better accuracy, lower inference times, and user-independent models. The use of huge and powerful processing units (GPUs, multicore servers, etc.) allows the use of deep learning approaches implementing very complex models (*Convolutional Neural Networks with several layers and neurons*) run in the cloud. Even, some commercial solutions allow local processing with the use of *specific microprocessors*, instead of cloud computing with remote servers working 24-day/7-hour and consuming very important amounts of energy. However, these solutions are devoted to non-critical applications, and when wearable systems are required with real-time and offline operation at the same time as low power consumption, cybersecurity, and data protection mechanisms, another approach is needed. In this scenario, simple artificial intelligence (AI) models are required to operate on the edge and share as little information as possible out of the edge. This simplification in AI modeling fits very well with the clustering-wise customization of models.

Furthermore, the research performed on the detection of emotions in people through the analysis of physiological and/or physical data has shown that there are very different emotional reactions in persons, depending on intrinsic and extrinsic factors [2]. Therefore, building a subject-independent model for emotion recognition is not the best solution. However, generating subject-dependent models makes the enrolment process and the application deployment almost unaffordable in most cases. In this dilemma, clustering persons according to similarities in

This work has been supported by the National Research Project Bindi-Tattoo PID2022-142710OB-I00 (Agencia Estatal de Investigación, AEI, Spain).

emotional responses would avoid subject-dependent models and would allow the enrollment of new users in the system. This clusterization will also allow to simplify AI models prototyped in the devices on the edge while sharing less amounts of personal data and releasing the wireless communication channels with unnecessary information.

In this work, personalized clustering is proposed to achieve an effective, accurate, and sustainable emotion detection system for applications running on the edge. The simpler AI algorithms run on the final devices reduce power consumption while increasing inference accuracy. The addition of new users is simple and cheap, and it does not require redesigning the whole system.

The original contributions of the papers are as follows:

- Introduction of a novel clustering methodology that aims to take advantage of similar physiological reactions by creating semi-customized models for each of the typologies of volunteers found in the data.
- Design of an extended version for the incorporation of new users in a simple way without the need to obtain data already labeled.
- Implementation of the complete data processing, clustering, model customization, and incorporation of a new volunteer pipeline.

The paper is organized as follows. Section II states the assumptions and basic elements of the emotion detection system and the current issues faced by researchers. Section III details the tools used in the research presented in this paper. Section IV presents the proposed methodology, and Section V shows the experimental results. Finally, in Section VI, the authors discuss the results and draw conclusions.

## II. Background.

The detection of emotions in humans using technological solutions has interested researchers in the last decades. The range of applications is widening every year. Starting with positive emotions identification, such as in marketing techniques to motivate potential shoppers or in entertainment and leisure activities to improve experiences, and going through negative emotions detection, in cases of mental health, surveillance or violence prevention, there have been several contributions in scientific research and commercial devices, especially in 21st century.

There is no global agreement on the range of emotions to be detected, neither their number nor their representation. However, the most accepted classifications are, for discrete emotions that are given by Ekman [3] with 12 main emotions (joy, sadness, surprise, contempt, hope, fear, attraction, disgust, tenderness, anger, calm, tedium), and for dimensional modeling, the Mehrabian proposal [4] of the PAD space (Pleasure, Arousal, Dominance) is frequently used in machine learning algorithms that prefer numerical values. Regarding the number of emotions, typically, a small number is preferred in multiuser applications.

Automatic emotion recognition is based on artificial intelligence algorithms that analyze one or multiple measurements from the person under analysis and classify the emotional reaction felt by her/him. This classification is done without the participation of the user, although subsequent confirmation is welcome to readjust the AI model. The key issue in Affective Computing is the type of variables measured from the person to infer the emotional state. Those variables can be physiological signals (Skin Temperature (SKT), Galvanic Skin Response (GSR), Heart Rate (HR), Electromyogram (EMG), Electrocardiogram (ECG), Electroencephalogram (EEG), etc.) caused for the instant reactions in the amygdala and sympathetic nervous system (SNS) and provoking the natural reaction of fight-or-scape. These signals should be captured through specific sensors. On the other hand, emotions can be reflected in human behavior and expressions, in face, voice, walking, etc. These are known as physical variables and can be captured through image or audio devices.

There has been intensive research about the best collection of variables to recognize emotional reactions with artificial intelligence. The scientists have finally agreed multimodal systems will always provide better results than unimodal ones. However, the number of sensors or measuring devices should be small so as the system could be worn easily, and the inference takes few seconds to provide the emotion detected. [5]

An emotion recognition system is built with three main elements: sensors, data, and algorithms. The first step is deciding the variables to be measured, which will imply the number and type of sensors to implement in the system. The second step is collecting data from these sensors with a real meaning, i.e., data measured from people while they are experiencing emotional reactions, together with the emotional label. These data will train the AI algorithm to generate a model able to classify emotions when receiving new data from other users. The system is characterized with different metrics that assess the level of accuracy in classifying the emotion. Depending on the data used for the training, and testing, as well as on the algorithms employed (KNN, SVM, NN, etc.), the quality of the system can vary from 60% to 80% of correct answers. The avoidance of bias in the data collection, a wide range of ages, both sexes, different and effective stimuli to provoke emotions, a good labeling system, etc. will produce good data sets to train and test the algorithms.

However, more problems appear when the system is deployed and applied to new users. The real-life conditions, external situations, multiple stimuli, different sensors, etc., will produce different measurements and, possibly, different emotional reactions, which will confuse the inference model. The generation of a subject-independent model appears in the literature as a bad idea, but subject-dependent models are very expensive in terms of time and resources. Clustering users' data, or other variables, seems a good solution [6].

There have been various works in the literature dealing with the use of clusters for AI algorithms focused on human reactions.

[7] propose an efficient method to improve the marketing techniques in e-commerce. Instead of clustering users by similarities, and recommending items bought by users in the same group, the solution proposed clusters items bought together and recommends users with items in the same cluster, according to the shopping cart. The addition of new elements in the clusters requires analyzing offline its purchasing history, while online recommendations for users take very few seconds. In this work, the data managed were users, items, and relations among these.

In [8], the authors propose the use of clusters to improve the learning process in students, based on the assessment question with lower rates of success. These clusters classify questions and, therefore, contents that allow the application of reinforcement schemes in a learning management system. The data processed were questions in the examination and assessments obtained. The clusters were created w.r.t. *question difficulty* and *item discrimination*, leveraging the number of students. The enrolment of new users (students) implied recalculating the clusters.

If we focus on emotion detection, the use of clustering has been growing in recent years. First, analyzing the utility of physiological signals in this detection, [9] propose the use of typologies to determine the best variables for recognizing emotional states (with a coarse classification: positive-negative or neutral-non neutral). In this work, the addition of new measurements is not considered. The data managed were physiological signals and emotion labels (target and self-reported).

Also, [10] apply clustering in emotion recognition through semantic analysis in conversations. The features extracted from the data collected are grouped into clusters, which are generated to reduce the high dimensionality. These clusters are located in a 3D PAD space (Pleasure-Arousal-Dominance), which states the position of the different emotions and their relative position in a more accurate layout. This change allows better and simpler AI models while although not every emotion is well detected.

More recently, [11] details the use of AI with KNN clustering for physiological variables (ECG and SKT) to generate a subject-independent model to classify the emotional state of intruders in secured environments. The addition of a new user in the system does not modify the original inference model. This scheme is proposed frequently in the literature. [12] classify emotions in students learning a foreign language with a multimodal approach (facial expression, audio, texts, and biological signals). With such many variables and features extracted, clusterization can be used to identify the most representative variables and generate a subject-independent inference model. The addition of a new user does not modify the original AI model.

There are some authors that treat the differences in emotion elicitation, considering the same external stimuli, not only labeling the emotion felt but also regarding the variation in their physiological variables. According to [2] previous experiences, post-traumatic stress disorders, personal traits, and current circumstances are conditioning greatly the emotional reaction and its length in time. These differences imply two main difficulties in automatic systems classifying emotions with AI algorithms. First, the dataset used for training and testing the algorithm should include those differences in emotional reactions, but if yes, a unique AI model will be very inaccurate. Previous clustering to identify similar emotional reactions would reduce this problem, in a midterm between subject-dependent and subject-independent AI models. Secondly, the addition of new users to the AI model should imply a previous analysis of which cluster he/she belongs to.

In [13] differences in emotional labeling are tackled. First, a subject-independent AI model is generated to classify the emotions felt by users. Secondly, those data with incorrect emotional classification are clustered and re-labeled, considering data with good classification results. Data used in this research are EEG signals from 11 and 32 channels. The addition of a new user is not considered.

Regarding the generation of personalized models, departing from many data for multiple users, we found two relevant works. In [14] only one AI emotion detection algorithm is generated, although the addition of new users is also considered. Physiological signals used are ECG, SKT, and GSR. Automatic feature calibration is generated for new users within the emotion detection system, previously to the classification model. Therefore, different sensors and/or conditions are overcome. Users' data are clustered with an unsupervised learning method, generating different centroids, which are used for comparison with the centroids of new users in the future. The best-correlated centroid will be used to calibrate the new data received and to recalculate the CNN, with the objective of a subject-independent model.

In [6] the real implementation of an emotion detection system is considered. Physiological data used are EEG, ECT, EMT, and GSR. Therefore, different users recruited in different moments, with different sensors and conditions, could be included in the emotion detection systems thanks to users' clustering and one AI model per cluster. Each cluster would produce a different number and type of features to generate its model. The experiments carried out for generating data from every user follow a similar scheme: baseline (no emotion), task (stress emotion), and recovery (no emotion). The data used for cluster generation are those from the "baseline" stage, therefore few differences are found among users, and only two clusters are generated. The rest of the data are used to generate the AI models. This, together with the small amount of data compiled (41 users, 11-11-19 30s-windows, 50% overlapped), makes it necessary to reconsider the approach.

Up to our knowledge, there is not an efficient methodology to include new users in emotion detection systems with a subject-dependent model based on clusters. In this work, we have proposed a solution that fits with sustainability, accuracy, and scalability.

### III. Tools

The following section provides detailed information concerning the tools necessary to implement the proposed methodology. The material includes the dataset, the machine learning model to test the clustering strategy influence, and the algorithm for computing the optimal number of clusters.

The complete pipeline and model implementation are developed using Python v3.8.10, leveraging its Scikit-learn and SciPy libraries.

#### A. UC3M4Safety

This research utilized the first release of The Women and Emotion Multi-modal Affective Computing (WEMAC) data [15], a multi-modal dataset integrated within the UC3M4Safety repository. The UC3M4Safety research team compiled and disseminated this dataset between 2020 and 2023. Its principal aim is to comprehend and formulate models delineating the relationship between physiological signals and fear activation [16].

This first set contains data from 47 women volunteers exposed to 14 validated audiovisual stimuli through a virtual reality environment. After each visualization, an interactive screen allows the participants to self-report the emotional labels. At the same time, their physiological information (galvanic skin response, skin temperature, and blood volume pulse signals) is collected by means of BioSignalsPlux [17].

This study is based on the direct utilization of 57 features extracted from physiological signals using 20-second windows with a 50% overlap, as outlined in previous research. A comprehensive description of the processing pipeline and its features can be found in (Miranda et al., 2021) [18].

Of the 47 volunteers, three were discarded for experimental procedures due to signal anomalies and lost data.

### B. K-Nearest neighbors' model (KNN)

The designed model to test the impact of profiling on fear classification performance is a binary supervised KNN machine-learning algorithm. It is an adapted version of the Scikit-learn model, incorporating the misclassification cost as a new hyperparameter. The model utilizes the Euclidean metric as the distance measure, with a misclassification cost of 1.6 for determining the class label.

Input features are standardized using an individualized Z-score normalization for each volunteer.

For the training, validation, and testing stages, a LOSO (Leave One Subject Out) strategy is proposed. This strategy involves reserving all the data of a single participant as the testing partition.

A stepwise Bayesian hyperparameter optimizer with five cross-validation folds is employed to determine the optimal hyperparameters. Model performance is evaluated using the accuracy and F1-Score.

The model's implementation is based on previously tested versions using the same dataset. Therefore, its performance is preliminarily validated by comparing it with results from prior studies conducted in Matlab®[18].

### C. Optimal cluster search algorithm

The search algorithm, guided by the Dunn index, identifies the optimal number of clusters (ranging from two to ten) by employing the hierarchical clustering method using the Euclidean distance metric and ward linkage criteria. Additionally, a prerequisite condition is imposed to ensure that each resulting cluster incorporates a minimum representative size of 15% of the total dataset. In the event that a cluster fails to meet this criterion, it is merged with the nearest cluster.

The Dunn index is a metric for evaluating clustering quality. It is derived from the ratio between the minimum inter-cluster distance and the maximum intra-cluster distance.

## IV. Proposed Methodology.

Two main methods have been designed and implemented to explore the potential of identifying typologies of reaction in volunteers and how these groupings can lead to personalized and improved models. In the first approach, M1, the volunteers are grouped into distinct categories based on statistical variables that correspond to the labeled observations of the two target classes. However, this technique's implementation requires the acquisition of multiple labeled observations from both target classes for the same participant.

The second approach, M2, is an extended version that brings the system as close as possible to a real use case. In this scenario, a volunteer with unlabeled observations can be assigned to one specific typology and take advantage of the customized model's benefits.

A. User profile clustering based on labeled data (M1)

This first technique aims to identify user typologies according to their emotional and physiological reactions.

The underlying concept is to detect analogous physiological responses to enhance data similarity, thereby facilitating the model in discerning between different classes. This approach enables a semi-personalized, yet broader model compared to a subject-dependent one.

For the initial grouping separation, an initial transformation ($\mathbb{R}^{OxF} \rightarrow \mathbb{R}^{NxM}$) of the feature map $D$ is required. Where the original dimensions $OxF$ (where $O$ represents the number of observations and $F$ represents the number of features) are converted into a matrix $D'$ of dimensions $NxM$ (where $N$ represents the number of volunteers and $M$ includes the mean and standard deviation of all the features for each of the two target classes: fear and non-fear) with every row a single entry for each volunteer. .

With this reshaped feature map and the algorithm previously presented, the optimal number of typologies and emotional similarities $K$ are found across all the training volunteers. Upon identification of these specific groups, the centroids $C_k \in \mathbb{R}^M$ for each cluster $k \in \{1, ..., K\}$ are computed.

Finally, new volunteers are integrated into the existing clusters by employing the minimum distance technique.

A brief scheme of the steps followed is presented in Figure 1.

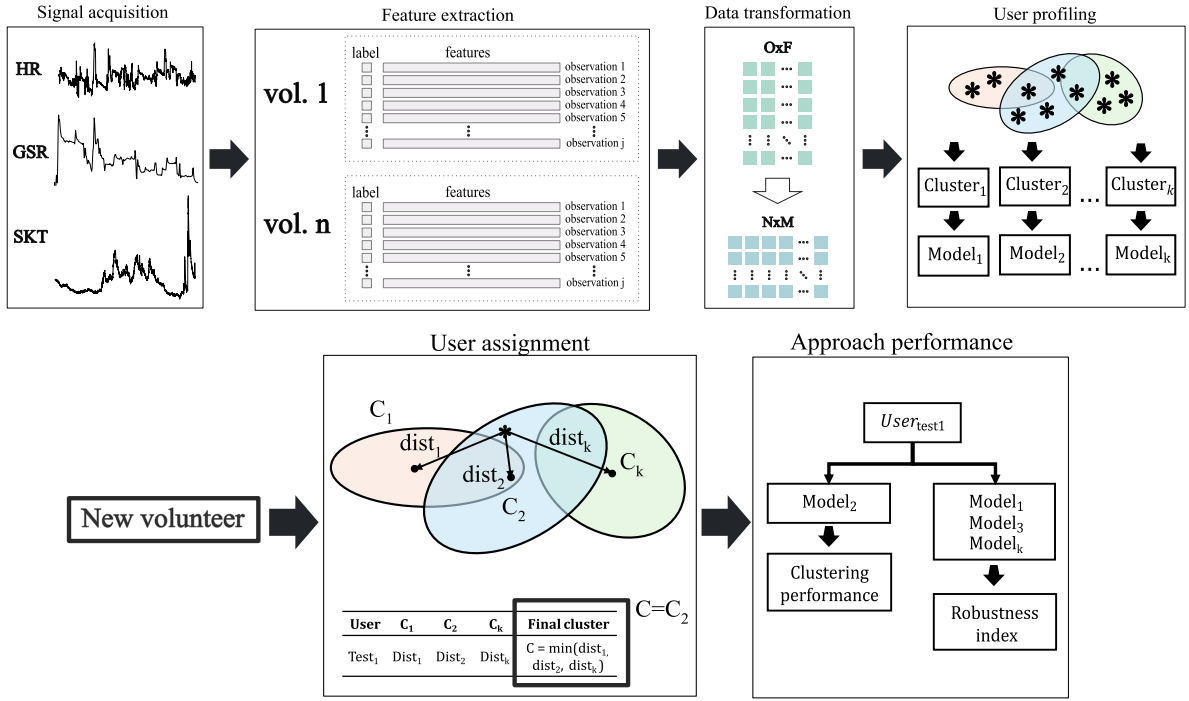

*Figure 1 User profile clustering based on labeled observations scheme*

B. Unlabeled observation clustering assignment – online (M2)

This approach aims to extend the proposed user clustering methodology to enable the assignment of any unlabeled observation to the closest profile, thereby creating a more complex but realistic technique. With this intention, the user typologies are considered the extension's initial starting point.

To aid understanding and maintain consistency, clusters found in M1 are referred to as typology clusters (TC), and the new clusters computed in this method are referred to as internal clusters (IC).

The primary challenge in assigning new observations, $O \in \mathbb{R}^F$, to an existing TC lies in the dimensional disparity between these two points. To mitigate this issue, the TCs centroids ($C_{tc} \in \mathbb{R}^M$) must be redefined to align with the dimensions corresponding to the features associated with each data instance, $\mathbb{R}^F$. This is going to be possible by means of the ICs.

To do this, all the training observations linked to each of these TCs are used to compute from four to six ICs that represent the behavior within each TC in terms of observations, ensuring that their centroids $C_{ic} \in \mathbb{R}^F$. These ICs are found using the algorithm to select the optimal number of groups.

Finally, when integrating a new volunteer into a specific TC, the cluster is selected based on the lowest summation of distances of the available observations. The data pipeline is presented in Figure 2.

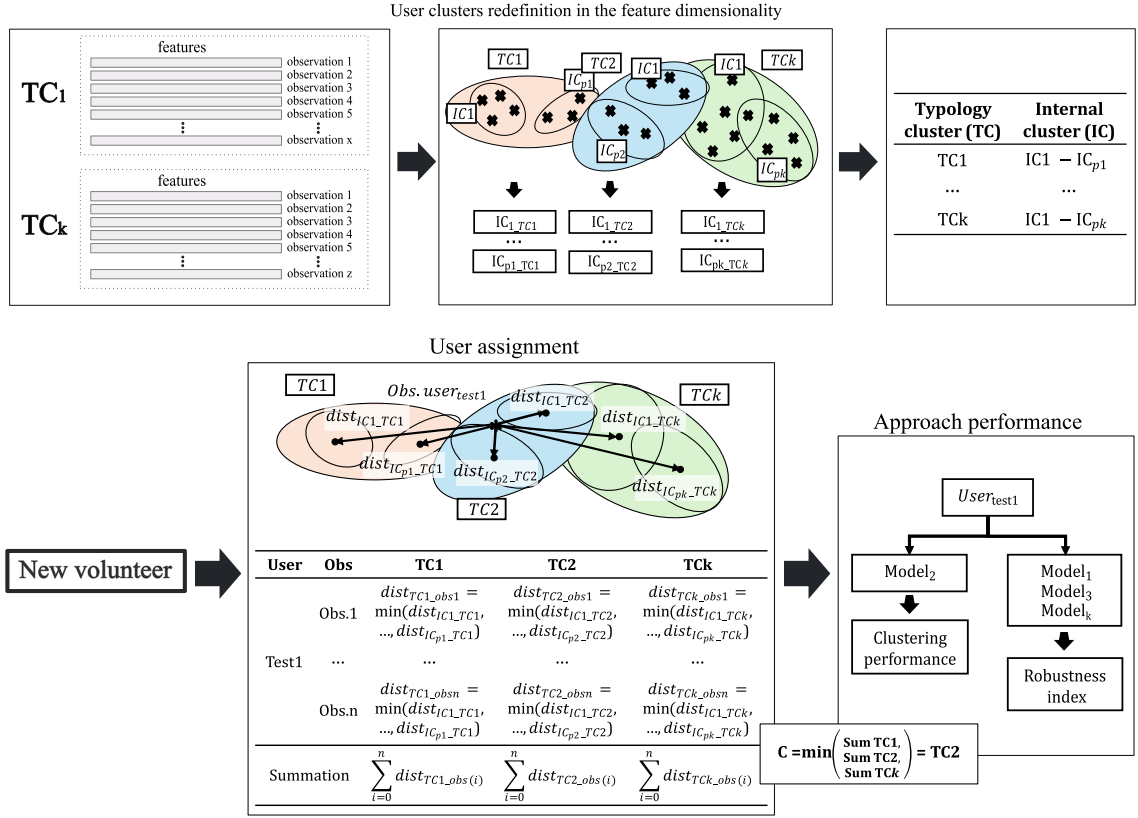

*Figure 2. Unlabeled observation clustering assignment scheme*

### V. Experimental Results.

This section details the experiments conducted to validate and test the proposed methodology and how it proceeds and influences the performance metrics of the KNN machine learning classifier.

Two main set-ups of experiments are shown and referred to: (1) Config.1 is the pipeline validation of the methodologies M1 and its extended version M2; (2) Config.2 presents the baseline, related works, and mean results analysis of the effectiveness of the semi-personalized methodology. Both experiments utilized the first release of the WEMAC dataset presented in Section III.

Config.1 evaluates the proposed methods M1 and M2. The experiment is divided into three validation indices: (1) Robustness test, where the model is tested with the volunteers that do not belong to that specific cluster; (2) Performance test, where the clustering model is tested with its assigned and unseen volunteers; (3) Volunteers assignment comparison test between methods.

In order to analyze the methodology, a 20-fold strategy is applied to generate a 70%-30 % training and testing partition. The training data set generates the four TCs following the M1 approach, and the criteria established in the search algorithm. The testing volunteers are assigned to an existing cluster using the same method.

With this clustering information, four different KNN models are trained: $model_k$ where $k \in \{1, ..., 4\}$.

Then, applying the same partitions and taking the TCs as a reference, the ICs are computed, and finally, the testing volunteers are assigned to one of the existing clusters using the method M2. The same validation procedure is repeated with these new assignments.

The validation results are displayed in Table 1. The improvements shown by cluster 1 are particularly noteworthy, almost reaching and even surpassing the 70% threshold in the performance metrics. Cluster 3 is the second with the best results, although it stays between 60 and 66%. The other two clusters show values of around 60%.

The second validation index, robustness, tries to prove the stability of clustering partitions. For this, the performance of personalized models using the volunteers assigned to the other cluster is computed to demonstrate that the values are clearly below the clustering model one. In other words, the volunteers of the rest of the clusters worsen the numbers because they belong to a different typology. In all cases, a difference in the performance of at least 5% is observed, except for the accuracy metrics of cluster 4, which presents almost the same levels for both tests. This shows that the typology represented in the fourth cluster is not very well defined and, therefore, works in a similar way with the volunteers assigned to this cluster and with the rest.

*Table 1. Average performance metrics (accuracy and f1-score) over 20 cross-validation folds for both methodologies and per typology cluster. Mean and standard deviation between brackets. The best results for each of the approaches are highlighted in bold.*

| Methodology | Typology Cluster (C) | Validation test | Accuracy | F1-Score |
|---|---|---|---|---|
| M1 | C1 | Robustness test | 53.11 (7.23) | 57.11 (6.45) |
| | | Clustering model$_1$ | **73.60 (6.58)** | **69.85 (6.31)** |
| | C2 | Robustness test | 50.10 (6.94) | 48.21 (10.05) |
| | | Clustering model$_2$ | 60.94 (7.02) | 58.84 (6.57) |
| | C3 | Robustness test | 57.57 (7.02) | 57.90 (7.37) |
| | | Clustering model$_3$ | 65.52 (7.61) | 63.02 (5.71) |
| | C4 | Robustness test | 58.73 (7.93) | 55.48 (11.91) |
| | | Clustering model$_4$ | 59.28 (6.23) | 63.38 (4.33) |
| M2 | C1 | Robustness test | 56.17 (7.23) | 59.24 (6.61) |
| | | Clustering model$_1$ | **70.91 (6.21)** | **68.24 (6.47)** |
| | C2 | Robustness test | 53.15 (8.08) | 49.49 (10.69) |
| | | Clustering model$_2$ | 59.94 (6.62) | 58.41 (6.86) |
| | C3 | Robustness test | 54.87 (11.22) | 53.64 (13.14) |
| | | Clustering model$_3$ | 62.15 (9.63) | 59.33 (4.51) |
| | C4 | Robustness test | 58.47 (6.31) | 54.19 (11.95) |
| | | Clustering model$_4$ | 58.68 (6.41) | 63.37 (4.18) |

Besides, analyzing the performance metrics and with the intention of verifying the distribution of the testing volunteers generated by both methods, a comparison between the M1 and M2 assignments to the same volunteer is made. A total of 84% of the total number of volunteers are assigned to the same cluster using both methods, which can be translated to similar results for both approaches.

The combination of all the previous results leads us to be able to say two things. The first is that both methods are equivalent in volunteer assignments; therefore, the use of the M2 has been validated despite its complexity. The second is that although not all groups show a clear performance improvement, in specific cases such as cluster 1, values of around 70% are reached.

Config.2 studies the mean results of the proposed techniques and compares them to our baseline and previous works. In this case, our three model configurations follow a training-testing LOSO strategy, and although, from the previous part, the equivalence between approaches has been established, both techniques are evaluated.

In Table 2, an increase of 3-4% and 1-2% in accuracy and f1-score metrics, respectively, are presented. Compared with our baseline, the semi-personalized models show a slight improvement. On the other hand, if compared with the related works, it is observed that in none of the cases their results are beaten. More specifically, the results obtained by the authors in [19] are well above those computed.

*Table 2. Comparison of the proposed methodology's performance metrics (accuracy and f1-score) with the related works. Mean and standard deviation between brackets.*

| Methodology | Accuracy | F1-Score |
| --- | --- | --- |
| Bindi [18] | 64.63 (16.56) | 66.67 (17.31) |
| Sun et al. [19] | 79.90 (04.16) | 78.13 (06.52) |
| General baseline model – without clustering | 60.75 (15.59) | 61.30 (12.49) |
| M1 clustering model – user profile clustering based on labeled data | 64.10 (06.80) | 63.30 (05.58) |
| M2 clustering model – unlabeled observations clustering assignment | 63.10 (07.10) | 62.90 (06.90) |

**Discussion and conclusions.**

This work proposes a novel clustering technique for semi-personalizing a general model and tackling the problem caused by the different physiological reactions to the same target emotion. This approach combines the initial utilization of user clustering based on labeled data to decide the existing typologies within the data, with the possibility of assigning new volunteers to their appropriate model later.

The experimental results, including the baseline comparison, robustness numbers, and semi-personalized results, show a slight improvement in the final performance values and a clear reduction of variation.

Although an improvement exists compared to the proposed general model, the performance metric values are not even close to those presented in [19] . This difference of more than 10% in both accuracy and f1-score makes it clear that migration from the KNN to a more complex deep learning model is necessary. With this new variation implemented, the proposed methodology should be retested to see how far it can be achieved and its actual effectiveness.

For that reason, future work will explore the use of more complex deep learning models like convolutional neural networks, or LSTMs, which have demonstrated significant effectiveness in recognizing spatial correlation and temporal patterns. But also, some optimizations for the assignment or clustering techniques.

Another main limitation of the study is the number of volunteers. There exists a possibility that the lack of data is preventing us from detecting other typologies that are currently agglomerated in the same cluster. For that reason, it would be interesting to perform this same study with a more extended data set, such as the second release of data from the same database used.